\documentclass[12pt]{article}%
\usepackage{amssymb}
\usepackage{amsmath}
\usepackage{amsfonts}
\usepackage{graphicx}%
\begin{document}

\begin{center}
\textbf{New classes of $n$-copy undistillable quantum states with negative
partial transposition}

Somshubhro Bandyopadhyay\footnote{Present address: Department of Chemistry, 80
Saint George Street, University of Toronto, Toronto, ON M5S 3H6, Canada
\par
Email: som@ee.ucla.edu} and Vwani Roychowdhury\footnote{Email:
vwani@ee.ucla.edu
\par
{}}

{\small Department of Electrical Engineering, UCLA, Los Angeles, CA 90095,
USA}

\end{center}

\begin{abstract}
{\small The discovery of entangled quantum states from which one cannot
distill pure entanglement constitutes a fundamental recent advance in the
field of quantum information. Such bipartite bound-entangled (BE) quantum
states \emph{could} fall into two distinct categories: (1) Inseparable states
with positive partial transposition (PPT), and (2) States with negative
partial transposition (NPT). While the existence of PPT BE states has been
confirmed, \emph{only one} class of \emph{conjectured} NPT BE states has been
discovered so far. We provide explicit constructions of a variety of
multi-copy undistillable NPT states, and conjecture that they constitute
families of NPT BE states. For example, we show that for every pure state of
Schmidt rank greater than or equal to three, one can construct $n$-copy
undistillable NPT states, for any $n\geq1$. The abundance of such conjectured
NPT BE states, we believe, considerably strengthens the notion that being NPT
is only a necessary condition for a state to be distillable. }

\end{abstract}

In the past decade, the search for efficient tools to determine whether a
given quantum state is entangled \cite{entanglement}, and if so, whether it
can be potentially used in quantum information processing protocols has led to
several fundamental results about the nature of quantum entanglement
\cite{peres, horo1, horo2, pawel, nielsen}. Almost all quantum communication
protocols, such as teleportation \cite{teleportation} and super-dense coding
\cite{sdc}, require maximally entangled states that are shared among the
spatially separated parties in conjunction with classical communication.
However, entangled states are noisy in general, due to environment induced
decoherence effects. Hence, in order for an entangled state to be useful, one
should be able to extract maximally entangled states (in the asymptotic sense)
starting from an ensemble of the given state, while using only local
operations and classical communication (LOCC). States which allow such
extraction of maximally entangled states are referred to as distillable
quantum states. Generalizations of classical information theory concepts have
led to protocols for distillation of quantum entanglement from certain classes
of quantum states \cite{popescu,deutsch,ibm1,gisin}.

Recent results have shown that even though most entangled states are
distillable, some are not. The undistillable but entangled quantum states are
said to possess bound entanglement \cite{pawel,peresbruss,horo3,ibm2}. Bound
entangled states cannot be prepared locally as they are entangled and being
not distillable, cannot be directly used in quantum communication protocols.
Interestingly, they can still enable non trivial quantum processes, such as
activation \cite{horo3, wolf} and superactivation of entanglement
\cite{smolin}, remote concentration of information \cite{murao}, that are not
possible with separable states. In this sense, the nonlocal properties of
bound entanglement can be distinctly utilized.

Many of the properties of entangled and distillable states are studied using
the partial transposition (PT) operation \cite{peres, horo1}. Let $\rho^{AB}$
be a density matrix corresponding to a bipartite quantum system consisting of
subsystems $A$ and $B.$ Then the partial transposition operation in an
orthogonal product basis is defined as: $\left(  \rho_{m\mu,n\nu}\right)
^{PT}=\rho_{m\nu,n\mu},$ where the transpose is taken with respect to the
subsystem $B.$ If $\rho^{PT}\geq0$, the state is said to be positive under PT
(PPT), otherwise it is said to be NPT. If a state is not entangled (i.e.,
separable) then it must be PPT. For quantum systems in $2\otimes2$ and
$2\otimes3$, the negativity under PT (NPT) is a necessary and sufficient for
inseparability but only sufficient in higher dimensions \cite{peres,horo1}. It
was proved that PPT states are not distillable
\cite{necsufcondfordistillation} and therefore, inseparable PPT states are
bound entangled. This leaves an interesting open question: Are all NPT states
distillable? If the answer is yes, then negativity under PT would be the
necessary and sufficient condition for distillability. However, it turns out
that even though most NPT states are distillable, some are possibly not. The
existence of NPT states that are not distillable has been conjectured in
\ Refs. \cite{NPTBE1,NPTBE2}.

A state $\rho$ is said to be distillable if and only if there exists a
positive integer $n$ and a Schmidt-rank (SR) two state $\left\vert
\phi\right\rangle $ such that \ $\left\langle \phi\right\vert \left(
\rho^{PT}\right)  ^{\otimes n}\left\vert \phi\right\rangle <0$
\cite{necsufcondfordistillation}. Intuitively this means that in the tensor
product Hilbert space $H^{\otimes n}$ there exists a $2\otimes2$ subspace
where the state is inseparable. Thus, a state is $n-copy$ \emph{undistillable}
if this condition is not satisfied by $n$ copies of the state. To prove that
NPT bound entangled states exist, one needs to show that the \emph{same state}
is $n$-copy undistillable for all $n\geq1$. In Refs \cite{NPTBE1,NPTBE2} the
conjectured NPT bound entangled states were proved to be one copy
undistillable. Moreover, for any given number of copies, $n$, states that are
$n$-copy undistillable were constructed. We explore this issue further, and
provide evidence that such conjectured NPT BE states can be found at
infinitely many neighborhoods of the Hilbert space.

\noindent\newline{\large \emph{General Approach}}\newline We now introduce a
general technique \cite{NPTBE1} to construct $n$-copy undistillable NPT
states. Consider a class of bipartite $d\times d$ density matrices
$\rho\left(  \epsilon\right)  $, $0\leq\epsilon\leq1$, where $\rho\left(
\epsilon\right)  $ is NPT when $\epsilon>0$, and PPT for $\epsilon=0$. Let us
also assume that \textit{the null space of the partial transpose of the
density matrix }$\left(  \rho\left(  \epsilon=0\right)  \right)  ^{\otimes n},
$\textit{for all } $d\geq3$\textit{\ and }$n\geq1$, \textit{\ does not contain
any non-zero vector of Schmidt rank less than three. }Now consider the
following function
\begin{equation}
f\left(  \epsilon,n\right)  =\min_{SR(\left\vert \phi\right\rangle
)=2}\left\langle \phi\right\vert \left(  \rho^{^{PT}}\left(  \epsilon\right)
\right)  ^{\otimes n}\left\vert \phi\right\rangle , \label{eq 1}%
\end{equation}
where, as shown, the minimum is taken over all Schmidt rank two states in the
full $d^{n}\otimes d^{n}$ Hilbert space. Since by assumption any state
$\left\vert \phi\right\rangle $ of Schmidt rank 2 does not lie in the null
space of $\left(  \rho^{\otimes n}\left(  \epsilon=0\right)  \right)  ^{PT}
$,\footnote{To be more precise, one needs to show that no Schmidt-Rank two
vector can lie arbitrarily close to the null space of $\left(  \rho^{\otimes
n}\left(  \epsilon=0\right)  \right)  ^{PT}$. However, it turns out that for a
finite $n$ it is sufficient to show that no Schmidt-Rank two vector lies in
the null space.} $f\left(  \epsilon=0,n\right)  >0$. Hence, it follows from
continuity arguments that for all $n$ there exists an $\epsilon_{n}$ such that
$f\left(  \epsilon_{n},n\right)  \geq0$, when $\epsilon$ is in the interval
$0\leq\epsilon\leq\epsilon_{n}$. This implies that there is a finite range of
$\epsilon>0$ for every $n$ such that $\rho\left(  \epsilon\right)  $ is
$n$-copy \emph{undistillable}. However, this argument is not sufficient to
conclude complete undistillability because we have not established any result
about the asymptotic behaviour of $\epsilon_{n}$ as $n\rightarrow\infty$. It
might so happen that $\epsilon_{n}\rightarrow0$ as $n\rightarrow\infty$ and
then we cannot guarantee the existence of a state that is undistillable for
any number of copies.

The purpose of the present note is to show that for quantum systems in
$d_{1}\otimes d_{2},d_{1},d_{2}\geq3,$ one can construct several classes of
states $\rho\left(  \epsilon\right)  $, $0\leq\epsilon\leq1$ such that
\textbf{(1)} $\rho\left(  \epsilon\right)  $ is NPT when $\epsilon>0$, and PPT
for $\epsilon=0$, and \textbf{(2)} the null space of the partial transpose of
the density matrix, $\left(  \rho\left(  \epsilon=0\right)  \right)  ^{\otimes
n}$, for any $n\geq1$, does not contain any nonzero vector of Schmidt rank
less than three. Then by arguments of the preceding paragraph, for any
$n\geq1$ we can generate states that are $n$-copy undistillable.

\noindent\newline{\large \emph{Constructions}}\newline\iffalse\fi We first
provide constructions of such $\rho\left(  \epsilon\right)  $ for a pair of
states $\left(  \sigma,\left\vert \varphi\right\rangle \left\langle
\varphi\right\vert \right)  $ that satisfy the following properties. \newline
Let $\sigma$ be an NPT state and let \ $\left\vert \varphi\right\rangle $ be a
pure entangled state of Schmidt rank $k$, $3\leq k\leq\min(d_{1},d_{2})$,
$\left\vert \varphi\right\rangle =\sum_{i=0}^{k-1}\lambda_{i}\left\vert
ii\right\rangle ,$ where $%
%TCIMACRO{\tsum \limits_{i=0}^{k-1}}%
%BeginExpansion
{\textstyle\sum\limits_{i=0}^{k-1}}
%EndExpansion
\left\vert \lambda_{i}\right\vert ^{2}=1$ and $\lambda_{i}$'s are real and
positive such that $\left\langle \varphi\left\vert \sigma^{PT}\right\vert
\varphi\right\rangle =-\left\vert \Lambda\right\vert $. \newline We will later
show that \emph{such combinations $\left(  \sigma,\varphi\right)  $ of states
are easy to construct}. In fact one can start with any arbitrary $\left\vert
\varphi\right\rangle $ and accordingly choose $\sigma$ and vice versa.

Very recently it has been shown that the operator $\frac{1}{D-1}\left(
\mathbf{I}-\left\vert \varphi\right\rangle \left\langle \varphi\right\vert
\right)  ^{PT}$where $\ \mathbf{I}$ is the identity operator of the total
Hilbert space $H$, is a separable density matrix \cite{bandyoroychowdhury}.
The proof that it is PPT is based on the eigen-decomposition of the partial
transposed operator $\left(  \left\vert \varphi\right\rangle \left\langle
\varphi\right\vert \right)  ^{PT}$:
\begin{equation}
\left(  \left\vert \varphi\right\rangle \left\langle \varphi\right\vert
\right)  ^{PT}=%
%TCIMACRO{\tsum \limits_{i=0}^{k-1}}%
%BeginExpansion
{\textstyle\sum\limits_{i=0}^{k-1}}
%EndExpansion
\lambda_{i}^{2}\left\vert ii\right\rangle \left\langle ii\right\vert +%
%TCIMACRO{\tsum \limits_{i,j=0,i<j}^{k-1}}%
%BeginExpansion
{\textstyle\sum\limits_{i,j=0,i<j}^{k-1}}
%EndExpansion
\lambda_{i}\lambda_{j}\left\vert \psi_{ij}^{+}\right\rangle \left\langle
\psi_{ij}^{+}\right\vert -%
%TCIMACRO{\tsum \limits_{i,j=0,i<j}^{k-1}}%
%BeginExpansion
{\textstyle\sum\limits_{i,j=0,i<j}^{k-1}}
%EndExpansion
\lambda_{i}\lambda_{j}\left\vert \psi_{ij}^{-}\right\rangle \left\langle
\psi_{ij}^{-}\right\vert , \label{eq 3}%
\end{equation}
where $\left\vert \psi_{ij}^{\pm}\right\rangle =\frac{1}{\sqrt{2}}\left(
\left\vert ij\right\rangle \pm\left\vert ji\right\rangle \right)  $.

We now construct the following density matrix
\begin{equation}
\rho\left(  \epsilon\right)  =\epsilon\sigma+\frac{\left(  1-\epsilon\right)
}{d^{2}-1}\left(  \mathbf{I}-\left\vert \varphi\right\rangle \left\langle
\varphi\right\vert \right)  ^{PT}. \label{Eq5}%
\end{equation}
It is easy to verify that the density matrix $\rho\left(  \epsilon\right)  $
is NPT when $\epsilon>0$, and separable for $\epsilon=0$: The negativity
follows from $\left\langle \varphi\right\vert \rho^{PT}\left(  \epsilon
\right)  \left\vert \varphi\right\rangle =\epsilon\left\langle \varphi
\left\vert \sigma^{PT}\right\vert \varphi\right\rangle =-\epsilon\left\vert
\Lambda\right\vert $.

The following result comprises the next crucial step in our proof for the
existence of $n$-copy undistillable states. \iffalse For a different class of
states the result was proved in Ref \cite{NPTBE1}. Although a part of our
proof follows a similar line of arguments, there are several important
differences, which we believe make the proof more transparent and understandable.\fi

\textbf{Lemma 1} Given a $\rho$ as defined in Eq.(\ref{Eq5}), the null space
of \textit{\ }$\left(  \rho^{PT}\left(  \epsilon=0\right)  \right)  ^{\otimes
n}$\textit{, }for all $d\geq3$ and $n\geq1$ does not contain any nonzero
vector of Schmidt rank less than three.

\textbf{Proof:} For a single copy the result is obvious since the only state
that lies in the null space of $\rho^{PT}\left(  \epsilon=0\right)  $ is
$\left\vert \varphi\right\rangle $ which of course has Schmidt rank greater
than two by construction. Before we outline our proof for n-copies, it is
instructive to work with two copies in detail because the proof contains all
the essential elements that we need for the case involving $n$-copies.

Let the following set be the basis for each of the Hilbert spaces
concerned:\newline%
\begin{equation}
\left[  \left\{  \left\vert \varphi_{l}\right\rangle \right\}  _{l=0}%
^{k-1},\left\{  \left\vert ij\right\rangle \right\}  _{i,j(i<j)=0}%
^{k-1},\left\{  \left\vert ij\right\rangle \right\}  _{i,j=k}^{d-1}\right]
\ , \label{eq 6}%
\end{equation}
where $\left\vert \varphi_{l}\right\rangle =%
%TCIMACRO{\tsum \limits_{i=0}^{k-1}}%
%BeginExpansion
{\textstyle\sum\limits_{i=0}^{k-1}}
%EndExpansion
\lambda_{il}\left\vert ii\right\rangle $. Note that $\left\langle \varphi
_{l}|\varphi_{s}\right\rangle =\delta_{ls}$ and furthermore, in this notation
$\left\vert \varphi_{0}\right\rangle =\left\vert \varphi\right\rangle .$ The
following set comprises a basis for the null space of the operator $\frac
{1}{\left(  d^{2}-1\right)  ^{2}}\left(  \mathbf{I}-\left\vert \varphi
\right\rangle \left\langle \varphi\right\vert \right)  ^{\otimes2}$:
\begin{equation}
\left[
\begin{array}
[c]{c}%
\left\vert \varphi_{0}^{1}\right\rangle \otimes\left\vert \varphi_{0}%
^{2}\right\rangle ,\left\vert \varphi_{0}^{1}\right\rangle \otimes\left\{
\left\vert \varphi_{l}^{2}\right\rangle \right\}  _{l=1}^{k-1},\left\vert
\varphi_{0}^{1}\right\rangle \otimes\left\{  \left\vert ij\right\rangle
^{2}\right\}  _{i,j(i<j)=0}^{k-1},\left\vert \varphi_{0}^{1}\right\rangle
\otimes\left\{  \left\vert ij\right\rangle \right\}  _{i,j=k}^{d-1},\\
\left\{  \left\vert \varphi_{l}^{1}\right\rangle \right\}  _{l=1}^{k-1}%
\otimes\left\vert \varphi_{0}^{2}\right\rangle ,\left\{  \left\vert
ij\right\rangle ^{1}\right\}  _{i,j(i<j)=0}^{k-1}\otimes\left\vert \varphi
_{0}^{2}\right\rangle ,\left\{  \left\vert ij\right\rangle ^{1}\right\}
_{i,j=k}^{d-1}\otimes\left\vert \varphi_{0}^{2}\right\rangle
\end{array}
\right]  \ . \label{eq 7}%
\end{equation}

Note that the superscripts indicate the individual Hilbert spaces. Let us
further simplify the notation before we proceed. We rewrite the above basis
as:
\begin{equation}
\left[  \left\vert \varphi_{0}^{1}\right\rangle \otimes\left\vert \varphi
_{0}^{2}\right\rangle ,\left\vert \varphi_{0}^{1}\right\rangle \otimes
\left\vert \varphi_{E}^{2}\right\rangle ,\left\vert \varphi_{0}^{1}%
\right\rangle \otimes\left\vert \varphi_{P}^{2}\right\rangle ,\left\vert
\varphi_{E}^{1}\right\rangle \otimes\left\vert \varphi_{0}^{2}\right\rangle
,\left\vert \varphi_{P}^{1}\right\rangle \otimes\left\vert \varphi_{0}%
^{2}\right\rangle \right]  , \label{eq 8}%
\end{equation}
where the sub-scripts $E$ and $P$ refer to entangled and product states
respectively. If there is a Schmidt rank two state in the null space it can be
written as a linear combination of the above basis states. Using the fact that
local projections cannot increase the Schmidt rank of a state, it readily
follows that the coefficients of the basis states that are of the form
$\left\vert \varphi_{0}^{1}\right\rangle \otimes\left\vert \varphi_{P}%
^{2}\right\rangle $ or $\left\vert \varphi_{P}^{1}\right\rangle \otimes
\left\vert \varphi_{0}^{2}\right\rangle $ are zero. If any of these
coefficients is not zero, then the reduced density matrix will have rank
$\geq3$. Therefore any Schmidt rank two state has to have the following form:
$\alpha\left\vert \varphi_{0}^{1}\right\rangle \otimes\left\vert \varphi
_{E}^{2}\right\rangle +\beta\left\vert \varphi_{E}^{1}\right\rangle
\otimes\left\vert \varphi_{0}^{2}\right\rangle $. It is useful to analyze this
explicitly. Let $\left\vert \psi\right\rangle $ \ be the Schmidt rank two
state and hence it can be written as,
\begin{equation}
\left\vert \psi\right\rangle =%
%TCIMACRO{\tsum \limits_{i=1}^{k-1}}%
%BeginExpansion
{\textstyle\sum\limits_{i=1}^{k-1}}
%EndExpansion
\alpha_{i}\left\vert \varphi_{0}^{1}\right\rangle \otimes\left\vert
\varphi_{iE}^{2}\right\rangle +%
%TCIMACRO{\tsum \limits_{i=1}^{k-1}}%
%BeginExpansion
{\textstyle\sum\limits_{i=1}^{k-1}}
%EndExpansion
\beta_{i}\left\vert \varphi_{iE}^{1}\right\rangle \otimes\left\vert
\varphi_{0}^{2}\right\rangle +\gamma\left\vert \varphi_{0}^{1}\right\rangle
\otimes\left\vert \varphi_{0}^{2}\right\rangle , \label{eq 9}%
\end{equation}
where the coefficients of the superposition are in general complex. On
substituting the expressions for the states and rearranging it in the
bipartite form one obtains
\begin{equation}%
%TCIMACRO{\tsum \limits_{j,l=0}^{k-1}}%
%BeginExpansion
{\textstyle\sum\limits_{j,l=0}^{k-1}}
%EndExpansion
\left(
%TCIMACRO{\tsum \limits_{i=1}^{k-1}}%
%BeginExpansion
{\textstyle\sum\limits_{i=1}^{k-1}}
%EndExpansion
\left\{  \alpha_{i}\lambda_{il}\lambda_{0j}+\beta_{i}\lambda_{0l} \lambda
_{ij}\right\}  +\gamma\lambda_{0j}\lambda_{0l}\right)  \left\vert
jl\right\rangle _{A}\left\vert jl\right\rangle _{B}. \label{eq 10}%
\end{equation}
The subscripts $A,B$ are used to emphasize the Schmidt form of the above
state. Note that (\ref{eq 10}) is already in a Schmidt decomposed form, where
the terms in the parentheses correspond to the Schmidt coefficients. If the
state is indeed of Schmidt rank two, then we must have all the coefficients
but two equal to zero. This amounts to solving $k^{2}$ linear equations for
$2k-1$ variables. One can explicitly write down the above equations in a
matrix form: $\mathbf{A}\mathbf{x}=\mathbf{y}$, where $\mathbf{A}$,
$\mathbf{x}$, and $\mathbf{y}$ are of dimensions $k^{2}\times(2k-1)$,
$(2k-1)\times1$, and $k^{2}\times1$, respectively ($k\geq3$). Moreover,
$\mathbf{y}$ has only two non-zero entries and $k^{2}-2$ zeros. The matrix
$\mathbf{A}$ can be shown to have the following property: Any submatrix of
$\mathbf{A}$ where \emph{any} two of the rows are deleted (hence, the
submatrix is of dimension $(k^{2}-2)\times(2k-1)$)) is still of full column
rank. Hence, $\mathbf{x}=\mathbf{0}$, and it would imply that the above state
has Schmidt rank zero. This completes the proof for two copies.

For $n$ copies, the proof follows the same line as for two copies. The basis
for the n-copy case of the null space is given by
\begin{equation}
\left\{  \left(
\begin{array}
[c]{c}%
n\\
m
\end{array}
\right)  \left\vert \varphi_{0}\right\rangle ^{\otimes m}\left\vert
\varphi_{l}\right\rangle ^{\otimes n-m}\right\}  ,m=0,...,n-1:l=1,..k-1
\label{eq 11}%
\end{equation}
Following the same arguments as in the two copy case, one can obtain a similar
set of linear equations. The number of equations is $k^{n}$ and the number of
variables can easily be counted and turns out to be $k^{n}-(k-1)^{n}$;
moreover, the right-hand-side of the equations (i.e., $\mathbf{y}$) has
$k^{n}-2$ zeros. Therefore, no matter how large $n$ may be, number of
equations is always greater than the number of variables, and one can show
from the properties of the matrix that the set of linear equations does not
have any non-trivial solution. \hfill$\square$

With the above result and the arguments provided in the beginning of the paper
we can now directly state the following theorem.

\textbf{Theorem 1 }\ Let $\sigma$ be a bipartite $d_{1}\times d_{2}$ (where
$d_{1},d_{2}\geq3$) NPT state and let $\left\vert \varphi\right\rangle $ be a
pure state of Schmidt rank equal to $k$ ($3\leq k\leq min(d_{1},d_{2})$), such
that $\left\langle \varphi\left\vert \sigma^{PT}\right\vert \varphi
\right\rangle =-\left\vert \Lambda\right\vert $. Then for any $n\geq1$, there
exists an $\epsilon_{n} >0$, such that the state
\begin{equation}
\rho\left(  \epsilon\right)  =\epsilon\sigma+\frac{\left(  1-\epsilon\right)
}{d^{2}-1}\left(  \mathbf{I}-\left\vert \varphi\right\rangle \left\langle
\varphi\right\vert \right)  ^{PT} \label{eq 12}%
\end{equation}
is \textit{$n$-copy undistillable} for $0<\epsilon\leq\epsilon_{n}$.

We now show that \emph{the pairs of states $\left(  \sigma,\varphi\right)  $
stipulated in Theorem 1 are fairly easy to construct}. In our first method we
will specify $\sigma$ first, and then accordingly we will specify $\left\vert
\varphi\right\rangle .$ In our second example, we will do just the opposite,
i.e., we will fix an arbitrary state $\left\vert \varphi\right\rangle $ and
based on the eigen decomposition of it's partial transpose, we will construct
$\sigma.$

\textit{Construction Method I:} Choose any pure entangled state $\left\vert
\psi\right\rangle $ of Schmidt rank $m$, $2\leq m\leq d-1$ of the form
$\left\vert \psi\right\rangle =%
%TCIMACRO{\tsum \limits_{i=0}^{m-1}}%
%BeginExpansion
{\textstyle\sum\limits_{i=0}^{m-1}}
%EndExpansion
\beta_{i}\left\vert ii\right\rangle ,$ where $\sum_{i=0}^{m-1}\left\vert
\beta_{i}\right\vert ^{2}=1$ and the Schmidt coefficients $\beta_{i}s$ are
real and positive. Let $\sigma=\left\vert \psi\right\rangle \left\langle
\psi\right\vert .$ Since the Schmidt rank of $\left\vert \psi\right\rangle $
is at most $(d-1),$ therefore there is at least one product state that is
orthogonal to the subspace spanned by the eigenvectors of $\sigma^{PT}.$ Let
$\left\vert \eta\right\rangle $ be such a product state. From the
eigendecomposition of any pure state (see Eq.(\ref{eq 3})) having Schmidt rank
greater than or equal to two, we know that the eigenvectors corresponding to
the negative eigenvalues are of Schmidt rank two. In particular, they are of
the form $\left\vert \psi_{ij}^{-}\right\rangle =\frac{1}{\sqrt{2}}\left(
\left\vert ij\right\rangle -\left\vert ji\right\rangle \right)  ,i<j$ with
negative eigenvalue $\beta_{i}\beta_{j}.$ Let $\left\vert \chi\right\rangle $
be one such eigenvector. Then, let $\left\vert \varphi\right\rangle
=\sqrt{\alpha}\left\vert \psi_{ij}^{-}\right\rangle +\sqrt{1-\alpha}\left\vert
\eta\right\rangle .$ For instance, if $\left\vert \psi\right\rangle $ has
Schmidt rank $(d-1)$, then $\left\vert \eta\right\rangle =\left\vert
dd\right\rangle .$ Clearly $\left\vert \varphi\right\rangle $ has Schmidt rank
three in this case but the Schmidt rank can be greater than $3$ if more than
one mutually biorthogonal product states that are also orthogonal to
$\sigma^{PT}$ can be found. This would be determined by the Schmidt rank of
$\left\vert \psi\right\rangle .$ It is now obvious that $\left\langle
\varphi\right\vert \sigma^{PT}$ $\left\vert \varphi\right\rangle =-\alpha
\beta_{i}\beta_{j}<0.$

\textit{Construction Method II: }\ Let us choose \emph{any arbitrary pure
state $\left\vert \varphi\right\rangle $ that has Schmidt rank }$k$, $3\leq
k\leq\min(d_{1},d_{2})$,
\begin{equation}
\left\vert \varphi\right\rangle =\sum_{i=0}^{k-1}\lambda_{i}\left\vert
ii\right\rangle , \label{eq 13}%
\end{equation}
where $%
%TCIMACRO{\tsum \limits_{i=0}^{k-1}}%
%BeginExpansion
{\textstyle\sum\limits_{i=0}^{k-1}}
%EndExpansion
\left\vert \lambda_{i}\right\vert ^{2}=1$ and $\lambda_{i}$'s are real and
positive. For any two operators $A$ and $B$ we have, $Tr\left(  AB^{PT}%
\right)  =Tr\left(  A^{PT}B\right)  $. For any $\sigma,$ we therefore have,
$\left\langle \varphi\left\vert \sigma^{PT}\right\vert \varphi\right\rangle
=Tr\left(  \left\vert \varphi\right\rangle \left\langle \varphi\right\vert
\sigma^{PT}\right)  =Tr\left(  \left\vert \varphi\right\rangle ^{PT}%
\left\langle \varphi\right\vert \sigma\right)  $. It follows from Eq.
(\ref{eq 3}) that if we choose $\sigma$ as the convex combination of the
eigenvectors with the negative eigenvalue of $\left\vert \varphi\right\rangle
^{PT}\left\langle \varphi\right\vert ,$then $\left\langle \varphi\left\vert
\sigma^{PT}\right\vert \varphi\right\rangle $ will be negative. We therefore
take the following representation of $\sigma$:%

\begin{equation}
\sigma=\sum_{i,j=0,i<j}^{k-1}\alpha_{ij}\left\vert \psi_{ij}^{-}\right\rangle
\left\langle \psi_{ij}^{-}\right\vert . \label{eq 14}%
\end{equation}
Then $\left\langle \varphi\left\vert \sigma^{PT}\right\vert \varphi
\right\rangle =-\sum\alpha_{ij}\lambda_{i}\lambda_{j}<0$. \iffalse The method
of our construction makes it clear that for every pure entangled $\left\vert
\varphi\right\rangle $ of Schmidt rank greater than or equal to three one
would be able to generate a finite copy undistillable NPT state. \fi

\noindent\newline{\large \emph{Generalized Constructions}}\newline We can
generalize our states in the following way. Let $m=\lfloor\frac{d}{k}\rfloor$,
where $\lfloor x\rfloor$ is the\textquotedblleft floor\textquotedblright%
\ operator denoting the largest integer less than or equal to $x$. Define the
following states: $\rho_{m}\left(  \epsilon\right)  =\epsilon\sigma
+\frac{\left(  1-\epsilon\right)  }{d^{2}-m}\left(  \mathbf{I}-%
%TCIMACRO{\tsum \limits_{i=1}^{m}}%
%BeginExpansion
{\textstyle\sum\limits_{i=1}^{m}}
%EndExpansion
\left\vert \varphi_{i}\right\rangle \left\langle \varphi_{i}\right\vert
\right)  ^{PT},$ where $\left\vert \varphi_{i}\right\rangle $ 's are pure
entangled states of Schmidt rank $k\geq3$ states such that each of them are in
orthogonal subspaces. Note that it is not necessary to have the Schmidt rank
of the states to be equal but the choice was made for simplicity and
convenience (notational). Clearly $m$ is maximum for a given $d$ when $k=3$.
The states are defined as follows $\left\vert \varphi_{i}\right\rangle =%
%TCIMACRO{\tsum \limits_{j=k(i-1)}^{ki-1}}%
%BeginExpansion
{\textstyle\sum\limits_{j=k(i-1)}^{ki-1}}
%EndExpansion
\lambda_{j}^{i}\left\vert jj\right\rangle ^{i}.$ As before, $\sigma$ may be
chosen to be the convex combination of the states with negative eigenvalues in
the eigen decomposition of the partial transpose of the pure states
$\left\vert \varphi_{i}\right\rangle .$ Note that $m=1$ corresponds to the
states in Theorem 1. One can then state the following generalization of
Theorem 1.

\textbf{Theorem 2}. The states
\begin{equation}
\rho_{m}\left(  \epsilon\right)  =\epsilon\sigma+\frac{\left(  1-\epsilon
\right)  }{d^{2}-m}\left(  \mathbf{I}-%
%TCIMACRO{\tsum \limits_{i=1}^{n}}%
%BeginExpansion
{\textstyle\sum\limits_{i=1}^{n}}
%EndExpansion
\left\vert \varphi_{i}\right\rangle \left\langle \varphi_{i}\right\vert
\right)  ^{PT} \label{eq 15}%
\end{equation}
for sufficiently small $\epsilon$ is\textit{\ n-copy undistillable} for any
$n\geq1$.

\noindent\newline{\large \emph{Distance from the Maximally Mixed State}
}\newline We next explore how these NPT $n$-copy undistillable states are
distributed in the Hilbert space, and in particular, how far they are from the
maximally mixed state. For any two quantum states $\rho_{1}$ and $\rho_{2}$,
the distance between the states is given by the Hilbert-Schmidt norm defined
by $\left\Vert \rho_{1}-\rho_{2}\right\Vert =\sqrt{Tr\left(  \rho_{1}-\rho
_{2}\right)  ^{2}}.$ Let us first note that the operator $\frac{1}{d^{2}%
-m}\left(  \mathbf{I}-%
%TCIMACRO{\tsum \limits_{i=1}^{n}}%
%BeginExpansion
{\textstyle\sum\limits_{i=1}^{n}}
%EndExpansion
\left\vert \varphi_{i}\right\rangle \left\langle \varphi_{i}\right\vert
\right)  ^{PT}$ is a PPT density matrix. The proof can be easily obtained by
using Eq. (\ref{eq 3}). Since our $n$-copy undistillable states exist
arbitrarily close to this state, it is sufficient to find the distance of this
state from the maximally mixed one. Using the H-S norm, one can show that the
distance is given by $\sqrt{\frac{m}{D(D-m)}}.$ We also note that this
distance is nothing but the distance of the maximally mixed state, $\frac
{1}{D}\mathbf{I}$, from any normalized $(D-m)$ dimensional projector
$\mathbf{I}_{D-m}$. Let us denote $r_{m}=$ $\left\Vert \frac{\mathbf{I}}%
{D}-\frac{\mathbf{I}_{D-m}}{D-m}\right\Vert .$

\textbf{Theorem 3 }\ For a bipartite quantum system in $d\otimes d,$ the
boundary of the balls of radius $r_{m}$ for all $m=1,..,\lfloor\frac{d}%
{k}\rfloor,k\geq3,$ around the maximally mixed state contains $n$-copy
undistillable NPT states. For a given $d,$ maximum number of such balls is
obtained when $k=3.$

It is instructive to analyze how close these states are relative to the
largest separable ball, the radius of which has recently been obtained in Ref.
\cite{gurvits}, and is given by $\frac{1}{\sqrt{D(D-1)}}.$ The result of
Theorem 3 shows that the case $m=1$ corresponds to the NPT $n$-copy
undistillable states that lie on the boundary of the largest separable ball.
This is as close as the states can be to the maximally mixed state. Let us now
try to answer how far from the maximally mixed state these NPT finite copy
undistillable states can be found. In our construction, for a given $d,$
maximum $r_{m}$ is obtained for $m=\lfloor\frac{d}{3}\rfloor.$ This
corresponds to a distance that grows as $\frac{1}{D^{1/4}}.$

\noindent\newline{\large \emph{Comparison with Previously Conjectured NPT BE
states} }\newline We now point out a remarkable similarity of the class of
states presented in this work with that obtained in Ref. \cite{NPTBE1}. It
turns out that for certain choices of the parameters in their class of states
and for a particular choice of $\left\vert \varphi\right\rangle $ in our case,
the null space of the partial transposed operator is exactly the same! Let us
denote the class of states in \cite{NPTBE1} as $\widetilde{\rho}\left(
c,\epsilon\right)  $ (following their notation). When $\epsilon=0$ and
$c=1/d(d+1),$
\begin{equation}
\left(  \widetilde{\rho}\left(  c=\frac{1}{d(d+1)},\epsilon=0\right)  \right)
^{PT}=\frac{1}{d^{2}-1}\left(
%TCIMACRO{\tsum \limits_{k=1}^{d-1}}%
%BeginExpansion
{\textstyle\sum\limits_{k=1}^{d-1}}
%EndExpansion
\left\vert \varphi_{k}\right\rangle \left\langle \varphi_{k}\right\vert +%
%TCIMACRO{\tsum \limits_{k,l=0,k\neq l}^{d-1}}%
%BeginExpansion
{\textstyle\sum\limits_{k,l=0,k\neq l}^{d-1}}
%EndExpansion
\left\vert kl\right\rangle \left\langle kl\right\vert \right)  \label{eq 16}%
\end{equation}
where, $\left\vert \varphi_{k}\right\rangle =\frac{1}{\sqrt{d}}%
%TCIMACRO{\tsum \limits_{j=1}^{d-1}}%
%BeginExpansion
{\textstyle\sum\limits_{j=1}^{d-1}}
%EndExpansion
e^{\frac{2\pi ijk}{d}}\left\vert jj\right\rangle ,k=1..,d-1.$ Going back to
our class, let us choose $\left\vert \varphi\right\rangle $ to be the
maximally entangled state of Schmidt rank $d$ (i.e., $\left\vert
\varphi\right\rangle =\frac{1}{\sqrt{d}}%
%TCIMACRO{\tsum \limits_{i=0}^{d-1}}%
%BeginExpansion
{\textstyle\sum\limits_{i=0}^{d-1}}
%EndExpansion
\left\vert ii\right\rangle $), \ instead of our general original choice of any
pure entangled state. If one expresses the identity operator as%
\begin{equation}
\frac{1}{d^{2}}\mathbf{I}=\frac{1}{d^{2}}\left(  \left\vert \varphi
\right\rangle \left\langle \varphi\right\vert +%
%TCIMACRO{\tsum \limits_{k=1}^{d-1}}%
%BeginExpansion
{\textstyle\sum\limits_{k=1}^{d-1}}
%EndExpansion
\left\vert \varphi_{k}\right\rangle \left\langle \varphi_{k}\right\vert +%
%TCIMACRO{\tsum \limits_{k,l=0,k\neq l}^{d-1}}%
%BeginExpansion
{\textstyle\sum\limits_{k,l=0,k\neq l}^{d-1}}
%EndExpansion
\left\vert kl\right\rangle \left\langle kl\right\vert \right)  \label{eq 17}%
\end{equation}
and substitutes Eq. (\ref{eq 14}) and $\left\vert \varphi\right\rangle $ in
that of $\rho^{PT}\left(  \epsilon=0\right)  $ (see Eq.~(\ref{eq 12}), one
obtains
\[
\rho^{PT}\left(  \epsilon=0\right)  =\frac{1}{d^{2}-1}\left(
%TCIMACRO{\tsum \limits_{k=1}^{d-1}}%
%BeginExpansion
{\textstyle\sum\limits_{k=1}^{d-1}}
%EndExpansion
\left\vert \varphi_{k}\right\rangle \left\langle \varphi_{k}\right\vert +%
%TCIMACRO{\tsum \limits_{k,l=0,k\neq l}^{d-1}}%
%BeginExpansion
{\textstyle\sum\limits_{k,l=0,k\neq l}^{d-1}}
%EndExpansion
\left\vert kl\right\rangle \left\langle kl\right\vert \right)  .
\]
The above similarity is striking considering the very different approaches
adopted in the two construction methodologies.

\noindent\newline\iffalse{\large \emph{Can one derive NPT BE states from our
approach?}}{\large \emph{Explicit Constructions of One-Copy Undistillable NPT
BE states} }\newline So far we have shown how to construct new classes of
$n$-copy undistillable NPT states based on analysis arguments. Staying within
our construction, we now obtain NPT states that we prove to be one-copy
undistillable. \ Let us choose $\sigma=$ $\left\vert \Psi^{-}\right\rangle
\left\langle \Psi^{-}\right\vert $ , where $\left\vert \Psi^{-}\right\rangle
=\frac{1}{\sqrt{2}}\left(  \left\vert 01\right\rangle -\left\vert
10\right\rangle \right)  $ is the familiar singlet state. The eigenvector
corresponding to the negative eigenvalue of $\sigma^{PT}$ is given by
$\left\vert \Phi^{+}\right\rangle =\frac{1}{\sqrt{2}}\left(  \left\vert
00\right\rangle +\left\vert 11\right\rangle \right)  .$ We now construct
$\left\vert \varphi\right\rangle $ as follows: $\left\vert \varphi
\right\rangle =\sqrt{1-\delta}\left\vert \Phi^{+}\right\rangle +\sqrt{\delta
}\left\vert 22\right\rangle $, where $\delta\ll1.$ From this construction it
is evident that the distillability of $\rho\left(  \epsilon\right)  $ implies
$\left\langle \Phi^{+}\right\vert \rho^{PT}\left(  \epsilon\right)  \left\vert
\Phi^{+}\right\rangle <0$, from which one easily obtains the constraint on
$\epsilon$ as a function of $\delta$ as given by $\epsilon>\frac{2\delta
}{d^{2}-1}$. This shows that for a given $\epsilon,$ one can always choose
$\delta$ such that $\epsilon\leq\frac{2\delta}{d^{2}-1}.$ Such states are
therefore provably one-copy undistillable.\fi

\noindent\newline{\large \emph{Discussions and Concluding Remarks} }\newline
We have shown that $n$-copy undistillable NPT states ($n\geq1$) exist at
infinitely many neighborhoods of the Hilbert space. Such states lie right on
the surface of the largest separable ball (LSB); thus, they are as noisy as
any inseparable state can be. They can also be found well outside of the LSB,
where distillable and separable states coexist. Can the general approach
adopted here lead to a proof of the existence of NPT BE states? Not in a
straightforward manner: In our constructions, $\rho^{PT}(\epsilon=0)$ has
$D-1$ identical nonzero eigenvalues, $\frac{1}{D-1}$. Hence, the function
$\displaystyle f\left(  \epsilon=0,n\right)  =\min_{SR(\left\langle
\phi\right\vert )=2}\left\langle \phi\right\vert \left(  \rho^{^{PT}}\left(
\epsilon=0\right)  \right)  ^{\otimes n}\left\vert \phi\right\rangle $ is
bounded above by $(\frac{1}{D-1})^{n}$, and $\displaystyle\lim_{n\rightarrow
\infty}f(\epsilon=0,n)=0$. Thus, we cannot claim that simple continuity
arguments will yield the existence of states that are NPT but undistillable
for any number of copies. However, since $\rho(\epsilon=0)$ is a separable
state (and hence, undistillable), one should expect $\displaystyle\lim
_{n\rightarrow\infty}f(\epsilon=0,n)=0$, and it provides no evidence that the
provably $n$-copy undistillable states do not remain undistillable for any
number of copies. In fact, \emph{ we conjecture that all the $n$-copy
undistillable states constructed here are also truly NPT bound entangled
states}. Moreover, we believe that even in our approach it is possible to show
that there exists a neighborhood $0\leq\epsilon\leq\epsilon_{\infty}$, where
$\displaystyle\lim_{n\rightarrow\infty}f(\epsilon,n)=0$; thus, proving that at
least all the states in this neighborhood are also NPT BE states.

\noindent\newline

{\large \emph{Acknowledgements}}\newline This work was sponsored in part by
the Defense Advanced Research Projects Agency (DARPA) project
MDA972-99-1-0017, and in part by the U. S. Army Research Office/DARPA under
contract/grant number DAAD 19-00-1-0172.

\end{document}